\documentclass[aps,prl,twocolumn,superscriptaddress,showpacs,preprintnumbers]{revtex4-1}

\usepackage{amsmath}
\usepackage{amsfonts}
\usepackage{amssymb}
\usepackage{txfonts}
\usepackage[dvipdfmx]{graphicx}
\usepackage{color}
\usepackage[colorlinks]{hyperref}
\usepackage{ulem}

\definecolor{CiteColor}{rgb}{0,0.5,0}
\hypersetup{citecolor=CiteColor}

\newcommand{\calE}{{\cal E}}
\newcommand{\calL}{{\cal L}}
\newcommand{\calO}{{\cal O}}
\newcommand{\atISCO}{_{\rm isco}}
\newcommand{\Hint}{{H_{\rm int}}}
\newcommand{\subgamma}{{\gamma}}

\newcommand{\rmd}{{\rm d}}

\definecolor{red  }{rgb}{1,0,0}
\definecolor{blue }{rgb}{0,0,1}
\definecolor{green}{rgb}{0,1,0}

\begin{document}

\title{Gravitational Self-Force Correction to the Innermost Stable Circular \\ Equatorial Orbit of a Kerr Black Hole}

\def\addYITP{Yukawa Institute for Theoretical Physics, Kyoto University, Kyoto, 606-8502, Japan}
\def\addNagoya{Division of Particle and Astrophysical Science,
Graduate School of Science, Nagoya University, Nagoya 464-8602, Japan}
\def\addKyoto{Department of Physics, Kyoto University, Kyoto 606-8502, Japan}
\def\addSoton{School of Mathematics, University of Southampton, Southampton SO17 1BJ, United Kingdom}
\def\addShef{Consortium for Fundamental Physics, School of Mathematics and Statistics, \\ University of Sheffield, Hicks Building, Hounsfield Road, Sheffield S3 7RH, United Kingdom}
\def\addLUTh{Laboratoire Univers et Th\'eories (LUTh), Observatoire de Paris, CNRS, \\ Universit\'e Paris Diderot, 5 place Jules Janssen, 92190 Meudon, France}
\def\addRoch{Center for Computational Relativity and Gravitation, Rochester Institute \\ of Technology, Rochester, New York 14623, USA}
\def\addUCD{School of Mathematical Sciences and Complex \& Adaptive Systems Laboratory, \\ University College Dublin, Belfield, Dublin 4, Ireland}

\author{Soichiro Isoyama}
\email{isoyama_at_yukawa.kyoto-u.ac.jp}
\affiliation{\addYITP}
\affiliation{\addNagoya}

\author{Leor Barack}
\affiliation{\addSoton}

\author{Sam R. Dolan}
\affiliation{\addShef}

\author{Alexandre Le Tiec}
\affiliation{\addLUTh}

\author{Hiroyuki Nakano}
\affiliation{\addKyoto}
\affiliation{\addRoch}

\author{Abhay G. Shah} 
\affiliation{\addSoton}

\author{Takahiro Tanaka}
\affiliation{\addYITP}
\affiliation{\addKyoto}

\author{Niels Warburton}
\affiliation{\addUCD}

\date{\today}


\preprint{KUNS-2490,YITP-14-33}


\begin{abstract}
For a self-gravitating particle of mass $\mu$ in orbit around a Kerr black hole of mass $M\gg\mu$, we compute the $\calO(\mu/M)$ shift in the frequency of the innermost stable circular equatorial orbit due to the conservative piece of the gravitational self-force acting on the particle. Our treatment is based on a Hamiltonian formulation of the dynamics in terms of geodesic motion in a certain locally defined effective smooth spacetime. We recover the same result using the so-called first law of binary black-hole mechanics. We give numerical results for the innermost stable circular equatorial orbit frequency shift as a function of the black hole's spin amplitude, and compare with predictions based on the post-Newtonian approximation and the effective one-body model. Our results provide an accurate strong-field benchmark for spin effects in the general-relativistic two-body problem. 
\end{abstract}

\pacs{04.25.Nx, 04.25.dg, 04.30.Db, 04.70.Bw}

\maketitle

{\it Introduction}.---A salient feature of orbital dynamics around Kerr black holes in general relativity is the existence of an {\it innermost stable circular orbit} (ISCO) for test particles. The ISCO radius depends on the magnitude of the spin of the black hole and on the orientation of the orbital plane with respect to the spin direction. Circular (timelike) geodesic orbits below the ISCO are unstable under perturbations away from circularity. The ISCO marks the onset of final merger in inspiraling compact-object binaries targeted by gravitational-wave detectors like KAGRA \cite{Somiya:2011np}, LIGO \cite{Harry:2010zz} and Virgo \cite{Accadia:2011}, as well as the future missions eLISA \cite{Seoane:2013qna} and DECIGO \cite{Kawamura:2011zz}.

The familiar notion of ISCO as a well-defined, precisely localizable marginal orbit is lost when the orbiting body is self-gravitating. For an object of small but nonzero mass $\mu\ll M$, where $M$ is the black hole's mass, the ISCO is replaced with a more vaguely defined {\it transition regime}, where the slow adiabatic inspiral, driven by radiation reaction, gradually transits into a direct plunge \cite{Ori:2000zn,Buonanno:2000ef}.

Nonetheless, a useful notion of ISCO can be retained even beyond the test-particle limit, if one focuses on the {\it conservative} dynamics of the binary system, ignoring dissipative effects. The value of the ISCO frequency for a nondissipating binary (of any mass ratio) is a useful diagnostic of the strong-field dynamics, and it has been playing an important role in the development of a general-relativistic two-body theory \cite{Damour:2000we,Blanchet:2001id,Damour:2009sm,LeTiec:2011dp,Favata:2010ic,Tessmer:2012xr,Pretorius:2005gq, Campanelli:2005dd, Baker:2005vv}. In the case $\eta:=\mu/M\ll 1$ the problem lends itself to perturbative methods. The orbiting object can be said to experience a gravitational self-force (GSF) \cite{Poisson:2011nh}, whose conservative piece causes an $\calO(\eta)$ ``shift'' in the location and frequency of the ISCO, relative to the test-particle case. This ISCO frequency shift is a valuable gauge-invariant characteristic of the strong-field dynamics beyond the geodesic approximation.  

Recent years have seen rapid progress in the development of rigorous methods for GSF calculations in black-hole spacetimes \cite{Barack:2009ux}. A milestone came in 2009 with the computation of the $\calO(\eta)$ ISCO frequency shift for a Schwarzschild black hole \cite{Barack:2009ey}. Many applications followed. For instance, the computed shift was used as a benchmark in an exhaustive survey of post-Newtonian (PN) methods and their performance in the strong-field regime \cite{Favata:2010yd}. It enabled the calibration of unknown parameters in the 
effective one-body (EOB) model \cite{Damour:2009sm}.
It also informed calculations of the remnant mass in astrophysical black-hole mergers \cite{Lousto:2009mf,Barausse:2012qz}, and provided crucial input for a recent model of intermediate-mass-ratio inspirals \cite{Lousto:2010qx}. These examples illustrate the usefulness of the ISCO frequency as a unique benchmark in general-relativistic dynamics. 

The value of the Schwarzschild ISCO frequency shift was computed in \cite{Barack:2009ey} based on a direct analysis of the restoring GSF effect on slightly eccentric orbits. This result was reproduced in later work \cite{LeTiec:2011dp,Akcay:2012ea}, with greater numerical accuracy,
using the first law of binary black-hole mechanics \cite{LeTiec:2011ab}. However, the important generalization to the Kerr case has not been tackled so far (except in a scalar-field toy model \cite{Warburton:2011hp}), primarily because GSF methods for Kerr spacetime are only now reaching maturity \cite{Dolan:2011dx,Shah:2012gu,Dolan:2012jg}.

In this Letter, we compute the $\calO(\eta)$ shift in the frequency of the ISCO in the equatorial plane of a Kerr black hole, as a function of the spin magnitude. Modeling spin effects in binary black-hole inspirals is a key priority in gravitational-wave physics, as black holes in nature are expected to carry significant spin \cite{Volonteri:2004cf,Reynolds:2013rva}. Our calculation sets an accurate strong-field benchmark for spin effects, and we expect it to provide a crucial input to this activity, in much the same way that the Schwarzschild result has impacted development in the field so far.  
 
Below we briefly describe our method and results; we relegate full details to a forthcoming paper \cite{Isoyama:2014-I}. Throughout we set $G = c = 1$, we use a metric signature $+2$, and we use $(t,\,r,\,\theta,\,\phi)$ to denote standard Boyer-Lindquist coordinates.

{\it ISCO in the test-mass approximation}.---To set the stage, we first review the formulation of ISCO for test particles, using the language of Hamiltonian mechanics. Consider a particle of mass $\mu$ and four-momentum $p_{\alpha}$, moving in the equatorial plane ($\theta=\pi/2$) of a Kerr black hole with mass $M \gg \mu$, spin $S =: a M =: q M^2$, and metric ${g}_{(0)\alpha\beta}$. Our convention is that $q>0$ ($q<0$) represents prograde (retrograde) orbits. Ignoring self-interaction, a Hamiltonian that generates geodesic motion is given by \cite{Carter:1968}
\begin{equation}\label{def-H0}
H_{(0)} (x^{\mu}, p_{\mu} ) :=
\frac{1}{2 \mu } {g}^{\alpha\beta}_{(0)}(x^{\mu})
\,{p}_\alpha {p}_\beta \, ,
\end{equation}
considered as a function on the 8D phase space spanned by the canonical variables $(x^{\mu}, p_{\mu})$. Hamilton's equations constrain the motion to a timelike geodesic of ${g}_{(0)\alpha\beta}$ with tangent four-velocity $u^\mu_{(0)} = g^{\mu\nu}_{(0)} p_\nu / \mu$, normalized as $g_{(0)\alpha\beta} u^{\alpha}_{(0)} u^{\beta}_{(0)} = -1$. The particle's energy $p_t =: - \mu \, \calE_{(0)}$ and angular momentum $p_\phi =: \mu \calL_{(0)} $ are constants of the motion.

Circular orbits satisfy $p_r = 0$ and $\dot{p}_r = 0$, where an overdot denotes a proper time derivative. The ISCO is identified by the vanishing of the restoring radial force $\dot{p}_{r}=-\partial H_{(0)}/\partial r$ under an arbitrary perturbation onto a slightly eccentric equatorial orbit. Since the variations of $\calE_{(0)}$ and $\calL_{(0)}$ are quadratic in the small eccentricity \cite{Barack:2009ey,Isoyama:2014-I}, and since (by definition) radial perturbations become stationary on the ISCO, it is sufficient to consider stationary perturbations with fixed $\calE_{(0)}$ and $\calL_{(0)}$ onto a nearby, nongeodesic circular equatorial orbit. This leads to the simple condition
\begin{equation}\label{ISCEO-G}
\left. 
\frac{\partial^2 H_{(0)}}{\partial r^2}\right\vert\atISCO = 0 \, .
\end{equation}
Together with $p_r = 0 $ and $\dot{p}_r = 0$, we have three equations for $\{r,\calE_{(0)},\calL_{(0)} \}$ at the ISCO location. One finds, in particular,
$
r^{(0)}\atISCO/ {M}=
3 + Z_2 \mp [(3 - Z_1) (3 + Z_1 + 2 Z_2)]^{1/2} ,
$ 
with 
$
Z_1 := 1 + (1 - q^2)^{1/3} [(1 + q)^{1/3} + (1 - q)^{1/3} ]
$ 
and 
$
Z_2 := (3 q^2 + Z_1^2)^{1/2}
$,
where the upper (lower) sign corresponds to prograde (retrograde) motion \cite{Bardeen:1972fi}.
The associated orbital frequency with respect to time $t$ reads
\begin{eqnarray}\label{Kerr-ISCEO}
M \Omega\atISCO^{(0)}
=
\left[ (r^{(0)}\atISCO / M)^{3/2} + q \right]^{-1} .
\end{eqnarray}

{\it ISCO in the perturbed spacetime}.---We now turn to examine $\calO(\eta)$ backreaction effects. It is well established~\cite{Mino:1996nk,Detweiler:2002mi,Pound:2009sm,Harte:2011ku} that, through $\calO(\eta)$, the particle follows a geodesic of a certain locally defined smooth effective geometry $g_{\alpha\beta}=g_{(0)\alpha\beta}+h^{R}_{\alpha\beta}$, where the second term ($\propto\eta$) is a particular solution to the linearized vacuum Einstein equation, obtained by subtracting a certain ``singular field'' $h_{\alpha\beta}^{S}$ from the physical, retarded metric perturbation $h_{\alpha\beta}^{\rm ret}$ associated with the orbiting particle.
Reference \cite{Isoyama:2013prep} develops a Hamiltonian formulation of the geodesic motion in $g_{\alpha\beta}$ based on 
the effective Hamiltonian
\begin{eqnarray}\label{Hamiltonian-G}
H[x^{\mu}, p_{\mu}; \gamma] = 
H_{\rm (0)} (x^{\mu}, p_{\mu} )
+ 
\Hint [x^{\mu}, p_{\mu}; \gamma] \,,  
\end{eqnarray}
where the first term is given by Eq.~\eqref{def-H0} and the interaction term reads (raising indices with the inverse background metric $g_{(0)}^{\alpha\beta}$)
\begin{equation}\label{H-int}
\Hint [x^{\mu}, p_{\mu};  \gamma]  :=
-\frac{1}{2 \mu } {h}^{\alpha\beta}_{R,{\rm sym}}
 [x^{\mu}; \gamma] 
\,{p}_{\alpha} {p}_{\beta}\,.
\end{equation}
Here, $\gamma$ is the ({\it a priori} unconstrained) trajectory that sources the metric perturbation and
$h_{\alpha\beta}^{R,{\rm sym}}:= \frac{1}{2} \, (h_{\alpha\beta}^{\rm ret}+ h_{\alpha\beta}^{\rm adv}) - h_{\alpha\beta}^{S}$
is the ``regularized'' time-symmetric part of the perturbation, responsible for the conservative piece of the GSF. 
We assume that a gauge is chosen such that $h^{R,{\rm sym}}_{\alpha\beta}$ manifestly respects the helical symmetry, and $h_{\alpha\beta}^{\rm {ret/adv}}$ vanish at infinity. This ensures that we can readily identify an ``asymptotic time'' coordinate $t$, for which the invariant orbital frequency (see below) can be defined unambiguously.
Hamilton's canonical equations for the effective Hamiltonian \eqref{Hamiltonian-G} read
\begin{eqnarray}\label{H-eq0}
u^\mu := \frac{\rmd x^{\mu}}{\rmd \tau} = 
\left.\frac{\partial H }
{\partial p_{\mu}}\right|_\gamma \, ,
\quad
\frac{\rmd p_{\mu}}{\rmd \tau} = -
\left.\frac{\partial H }
{\partial x^{\mu}}\right|_\gamma \, ,
\end{eqnarray}
where the proper time $\tau$ is measured along $\gamma$ in $g_{\alpha\beta}$. We find $u^\mu = g^{\mu\nu} p_\nu / \mu$, with the normalization $g_{\alpha\beta} u^{\alpha} u^{\beta} = -1$.

For circular orbits ($p_r = 0$ and $\rmd p_r/\rmd \tau = 0$), 
it follows from Eq.~\eqref{H-eq0} and the symmetry of $H_{\rm int}$ 
under Mino's transformation
~\cite{Isoyama:2013prep, Mino:2003yg, Hinderer:2008dm} that 
$\calE:=-u_t$ and $\calL:=u_\phi$ are constant along $\gamma$. 
Restricting to circular orbits, we write $H=H[\zeta^I;\gamma(\zeta_{\gamma}^I)]$, where we make a distinction between the phase-space coordinates $\zeta^I:=\{r,\, \calE,\, \calL \}$ and the parameters $\zeta^I_{\gamma}:=\{r_{\gamma},\, \calE_{\gamma},\, \calL_{\gamma} \}$ labeling the source trajectory. 
Now, in analogy with Eq.~\eqref{ISCEO-G}, the ISCO is identified by requiring the stationarity of the restoring force with respect to radial perturbations with fixed $\calE,\calL$ and also fixed $\calE_{\gamma},\calL_{\gamma}$. 
Here, however, the restoring force, $-\partial H/\partial r$, depends separately on both $r$ and  $r_{\gamma}$, which are treated as two independent degrees of freedom at the Hamiltonian level. We must therefore make sure to vary simultaneously the radial positions of both the orbit and the source before identifying one with the other. Hence, 
the stationarity condition reads
\begin{equation}\label{ISCOcond-r}
\left. \left(\frac{\partial}{\partial r}
 + \frac{\partial}{\partial r_{\subgamma}}\right)
   \frac{\partial H[\zeta^I;\gamma(\zeta_{\gamma}^I)]}
   {\partial r}\right\vert\atISCO = 0 \,. 
\end{equation} 
Reference \cite{Buonanno:2002ft} obtained a similar ISCO condition 
in the PN context.

{\it ISCO shift in terms of redshift variable}.---Following \cite{Detweiler:2008ft}, and restricting to circular orbits, let us introduce the ``redshift'' function
$
z := 1/{u^t}.
$
From 
${g}_{\alpha\beta} {u}^{\alpha} {u}^{\beta} = - 1$
follows 
\begin{equation}\label{z-identity}
z = \calE - \Omega \calL  \, ,
\end{equation}
where $\Omega = {u^{\phi}}/{u^t} = {\rmd \phi}/{\rmd t}$ is the circular-orbit frequency. Let us formally expand 
$z(\Omega) = z_{(0)}(\Omega) + \eta \, z_{(1)}(\Omega) + \calO(\eta^2)$,
where $z_{(0)}(\Omega)$ is the functional relationship for a Kerr geodesic, and $\eta \, z_{(1)}(\Omega)$ is the leading GSF correction for a fixed $\Omega$. We similarly expand $\calE$ and $\calL$. Varying $g_{\alpha\beta}u^{\alpha}u^{\beta}=-1$ with respect to $\eta$ at fixed $\Omega$ then gives
$\calE_{(1)} - \Omega \calL_{(1)} 
= z_{(0)} \Hint / \mu$, 
whereupon Eq.~\eqref{z-identity} becomes 
$z = z_{(0)} \, (1 + \Hint / \mu)$.
We thus obtain a simple link between $\Hint(\Omega)$ and the $\calO(\eta)$ piece of $z(\Omega)$, namely
\begin{equation}\label{def-z1}
\frac{\Hint}{\mu} = \eta \, \frac{z_{(1)}}{z_{(0)}}\,.
\end{equation}

Using Eq.~\eqref{def-z1}, one can show that Eq.~\eqref{ISCOcond-r} reduces to the remarkably simple condition 
\begin{equation}\label{ISCOcond-W}
{\tilde z}''(\Omega\atISCO) = 0 \,,
\end{equation}
where a prime denotes $\rmd / \rmd \Omega$, and we introduced the modified redshift function
$
{\tilde z}(\Omega) := z_{(0)}(\Omega) + \frac{1}{2} \, \eta \, z_{(1)}(\Omega) + \calO(\eta^2) = z(\Omega) [ 1- \Hint(\Omega) / (2\mu ) ] 
+ \calO(\eta^2) 
$.
We now sketch the derivation of this result, central to our analysis;
details will be given in~\cite{Isoyama:2014-I}.
The basic idea is to apply
$(\rmd / \rmd \Omega)
= r'(\partial / \partial r)
+ \calE'(\partial / \partial \calE)
+ \calL'(\partial / \partial \calL)
$ to the identity $H = -\frac{1}{2}\mu$. 
From $\rmd H / \rmd \Omega=0$, we obtain
$\calE'- \Omega \calL'= z H'_{\rm int} / (2 \mu ) $
because $\rmd {p}_{r} / \rmd \tau  = -\partial H / \partial r = 0$
along circular orbits. We also have $ 
(\rmd / \rmd \Omega) \partial H / \partial r = 0$, which implies
$(u^t)' \calE' - (u^\phi)' \calL' = {\cal O} (\eta^2)$
at the ISCO with the aid of Eq.~\eqref{ISCOcond-r}.
These two relations are combined to give ${\calE}'_{(0)}(\Omega^{(0)}\atISCO)
={\cal L}'_{(0)}(\Omega^{(0)}\atISCO) = 0$, and 
are rewritten as
$z'=-{\calL}+\frac{1}{2} ( z / \mu ) H'_{\rm int}$ and
$z'(z'+{\calL}) +z{\calL}'= {\cal O} (\eta^2)$.
From here, simple algebra gives $\tilde z''(\Omega\atISCO)={\cal O} (\eta^2)$.
This establishes the equivalence of Eqs.~\eqref{ISCOcond-r} 
and~\eqref{ISCOcond-W}. 

Following Refs.~\cite{Damour:2009sm,Favata:2010yd,LeTiec:2011dp,Akcay:2012ea}, we parametrize the ISCO frequency shift due to the conservative GSF in the form 
\begin{equation}
\label{def-ISCO}
(M + \mu) \, \Omega\atISCO
:= M \Omega\atISCO^{(0)} (q)
\left\{ 1 + \eta \, C_\Omega(q) + \calO(\eta^2) \right\} .
\end{equation}
Substituting in Eq.~\eqref{ISCOcond-W}, expanding through $\calO(\eta)$, and using the ISCO condition $z''_{(0)}(\Omega^{(0)}\atISCO) = 0$ for the background Kerr geodesic, where
$z_{(0)}^2 =
( 1 - a \Omega )
\bigl[ 1 + a \Omega 
-3 (M \Omega)^{2/3} 
(1 - a \Omega)^{1/3} \bigr]
$, we find
\begin{equation}\label{ISCO-shift}
C_\Omega = 1 - \frac{1}{2} \,
\frac{z''_{(1)}(\Omega^{(0)}\atISCO)}
{{\Omega}^{(0)}\atISCO \, z'''_{(0)}(\Omega^{(0)}\atISCO)}\,.
\end{equation}
This is our main formal result. It has a convenient form, 
as it involves only the redshift function along equatorial {\it circular} 
orbits.

{\it First law of binary mechanics}.---Before proceeding to numerical implementation, we show that Eq.~\eqref{ISCO-shift} is recoverable using the notion of minimum-energy circular orbit (MECO), within the framework of the first law of binary black-hole mechanics \cite{Friedman:2001pf,LeTiec:2011ab,Blanchet:2012at,Tiec:2013kua}. Starting from the perturbed Hamiltonian in Eq.~\eqref{Hamiltonian-G}, one can derive a first-law-type relationship that extends to $\calO(\eta)$ the test-mass results of \cite{Tiec:2013kua}. This variational relation holds for generic bound orbits.
For circular orbits, it reduces to (discarding changes in the black-hole mass and spin) \cite{Isoyama:2013prep} 
\begin{equation}\label{1st-law}
	\delta E = \Omega \, \delta L + z \, \delta \mu \, ,
\end{equation}
where we defined the binding energy 
$E := \mu \calE - \frac{1}{2} H_\text{int} \, \calE_{(0)} $ 
and the angular momentum 
$L := \mu \calL - \frac{1}{2} H_\text{int} \, \calL_{(0)}$. 
Notice that the combination ${\cal M} = E - \Omega L$, which can heuristically be viewed as the binding energy in a corotating frame, coincides with the modified redshift function: ${\cal M} = \mu \, \tilde{z}$.

The first law,~\eqref{1st-law}, implies the partial differential equations $\partial E / \partial \Omega = \Omega \, \partial L / \partial \Omega$ and $\partial E / \partial \mu = \Omega \, \partial L / \partial \mu + z$, which can be combined to give $E = {\cal M} - \Omega \, \partial {\cal M} / \partial \Omega$, $L = - \partial {\cal M} / \partial \Omega$ and $z = \partial {\cal M} / \partial \mu$.
We expand 
$E = \mu \bigl[ E_{(0)}(\Omega) + \eta \, E_{(1)}(\Omega) + \calO(\eta^2) \bigr]$, and similarly for $L$ and ${\cal M}$. Varying these equations 
with respect to $\eta$ at fixed $\Omega$ then gives $E_{(1)} = {\cal M}_{(1)} - \Omega {\cal M}'_{(1)}$, $L_{(1)} = - {\cal M}'_{(1)}$, and $z_{(1)} = 2 {\cal M}_{(1)}$. Eliminating ${\cal M}_{(1)}$, we obtain
\begin{equation}\label{E-GSF}
E_{(1)} = \frac{1}{2} \left( z_{(1)} - \Omega \, z'_{(1)} \right) .
\end{equation}

On the other hand, by definition the MECO minimizes the binding energy $E(\Omega)$. Thus, its frequency $\Omega_{\rm meco}$ is a solution of
\begin{equation}\label{def-MECO}
E'(\Omega_{\rm meco}) = 0 \, .
\end{equation}
For Hamiltonian systems such as ours, $\Omega_{\rm meco} = \Omega\atISCO$  
(cf. Ref. \cite{Buonanno:2002ft}).
Hence, using Eqs.~\eqref{def-ISCO} and \eqref{E-GSF} together with the test-particle relation $E'_{(0)}(\Omega^{(0)}_{\rm meco}) = 0$, Eq.~\eqref{def-MECO} yields
\begin{equation}\label{MECO-shift}
C_\Omega 
= 1 + \frac{1}{2} \, \frac{z''_{(1)}(\Omega^{(0)}\atISCO)}{E''_{(0)}(\Omega^{(0)}\atISCO)} \,.
\end{equation}
Equation \eqref{ISCO-shift} is then easily recovered by using $E''_{(0)} = - \Omega \, z'''_{(0)}$ at $\Omega = \Omega^{(0)}\atISCO$, obtained by noting that ${\cal M}_{(0)} = \mu \, z_{(0)}$ and hence $E_{(0)} = z_{(0)} - \Omega \, z'_{(0)}$, from which also follows $z''_{(0)}(\Omega^{(0)}\atISCO)=0$. Indeed, since ${\cal M} = \mu \, \tilde{z}$, Eq.~\eqref{ISCOcond-W} is equivalent to the MECO condition \eqref{def-MECO}.

{\it Numerical implementation and results}.---The evaluation of $C_{\Omega}(q)$ via Eq.~\eqref{ISCO-shift} requires as input numerical data for $z''_{(1)}(\Omega^{(0)}\atISCO)$. Reference \cite{Shah:2012gu} prescribes the construction of $z_{(1)}$ from the field $h_{\alpha\beta}^{R,{\rm sym}}$ for given $q$ and $\Omega$. The perturbation $h_{\alpha\beta}^{R,{\rm sym}}$ itself is obtained by applying a suitable regularization procedure to the (numerically computed) retarded metric perturbation sourced by a particle on a circular equatorial geodesic orbit of angular frequency $\Omega$. One then obtains a numerical representation of the function $z_{(1)}(\Omega)$, from which $z''_{(1)}(\Omega^{(0)}\atISCO)$ is extracted. This is repeated for a sample of $q$ values.

To obtain $C_{\Omega}(q)$, we used two independent numerical codes that are based on different numerical and regularization methods. In method (i), due to Shah {\it et al.}~\cite{Shah:2012gu}, the metric perturbation is reconstructed in a radiation gauge from frequency-domain numerical solutions to the Teukolsky equation \cite{Teukolsky:1972}, followed by an application of (a variant of) the standard mode-sum regularization technique \cite{Barack:1999wf}. Method (ii), 
developed in Refs.~\cite{Dolan:2011dx, Dolan:2012jg,Heffernan:2012vj,Dolan:2014prep}, 
uses a direct time-domain implementation of the linearized Einstein equation in Lorenz gauge
 (apart from a simple gauge transformation to assure asymptotic flatness \cite{Sago:2008id,Damour:2009sm}), and applies $m$-mode regularization~\cite{Barack:2007we}. 
 
For each $q$, we computed $z_{(1)}$ at orbital radii $r^{(0)} =  r^{(0)}\atISCO + \Delta r$ with $\Delta r / (0.05M) = -n,  \ldots, n$ where $n = 10$ and $n=4$ in cases (i) and (ii)~\cite{PRL-SM}.
We found the first and second derivatives of $z_{(1)}$ with respect to $r^{(0)}$ at $\Delta r = 0$ by fitting each data set with a polynomial in $\Delta r / M$. This gave $z_{(1)}''$ and thus $C_{\Omega}$ via Eq.~(\ref{ISCO-shift}). 

\begin{figure}[tbp]
\includegraphics[width=\columnwidth]{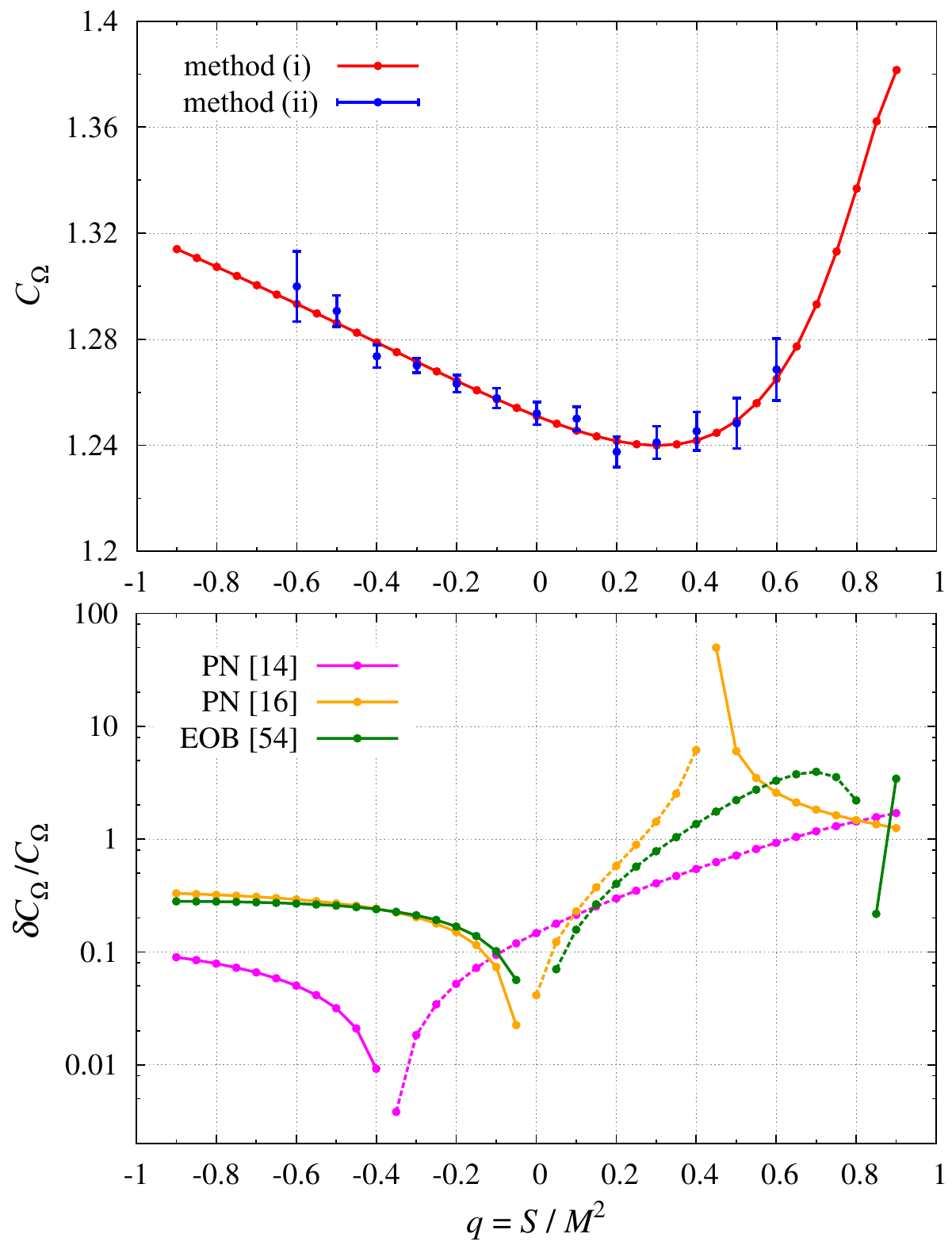}
\caption{
{\it Upper panel:} Relative ISCO frequency shift $C_\Omega$ [see Eq.~\eqref{def-ISCO}] as a function of the dimensionless black-hole spin $q = S/M^2$. The red line interpolates the accurate numerical results from our method (i), while blue data points and error bars are from our method (ii). {\it Lower panel:} Relative difference
$\delta C_{\Omega} / C_{\Omega} := 1 - C^{\rm PN/EOB}_\Omega / C_{\Omega}$ between our ``exact'' GSF results and earlier predicted values for $C_{\Omega}(q)$ from PN approximations \cite{Favata:2010ic,Tessmer:2012xr} and the EOB model of Ref.~\cite{Barausse:2009xi}, the latter having been calibrated so as to reproduce the correct value for $C_{\Omega}(0)$. The solid (dashed) lines indicate $\delta C_{\Omega} > 0$ ($\delta C_{\Omega} < 0$).
}
\label{fig:ISCEO-shift}
\end{figure}

Method (i) delivers highly accurate results, and method (ii) provides important checks. We found consistency between the data sets for $z_{(1)}$, and the derived values of $C_{\Omega}$, roughly to within the error bars of method (ii).
We note for method (ii) that (1) the dominant source of error for $|q| \gtrsim 0.5$ arises from a frequency-filter method for eliminating a linear-in-time gauge mode in the dipole sector \cite{Dolan:2012jg}, and 
(2) the error bars in Fig.~\ref{fig:ISCEO-shift} are estimated using a robust combination of Monte Carlo and bootstrap methods.
[No comparison data are available for $|q| > 0.6$, where method (ii), in its current state of development, becomes ineffectual.]


\begin{table}
\begin{tabular}{clc|ccl}
\toprule
$q$ & \quad\, $C_{\Omega}$ & & & $q$ & \quad\, $C_{\Omega}$ \\
\hline
$0.1$ & $1.245\,568(2)$ &&& $-0.1$ & $1.257\,379(3)$ \\
$0.2$ & $1.241\,595(4)$ &&& $-0.2$ & $1.264\,284(4)$ \\
$0.3$ & $1.239\,927(4)$ &&& $-0.3$ & $1.271\,478(7)$ \\
$0.4$ & $1.241\,83(1)$  &&& $-0.4$ & $1.278\,787(6)$ \\
$0.5$ & $1.249\,234(6)$ &&& $-0.5$ & $1.286\,093(9)$ \\
$0.6$ & $1.265\,030(8)$ &&& $-0.6$ & $1.293\,314(9)$ \\
$0.7$ & $1.293\,19(1)$  &&& $-0.7$ & $1.300\,397(7)$ \\
$0.8$ & $1.336\,83(2)$  &&& $-0.8$ & $1.307\,305(8)$ \\
$0.9$ & $1.381\,57(3)$  &&& $-0.9$ & $1.314\,016(9)$ \\
\botrule
\end{tabular}
\caption{Relative ISCO frequency shift $C_\Omega$ [see Eq.\ (\ref{def-ISCO})] for a sample of values of the spin parameter $q = S/M^2$. Negative $q$ values denote retrograde orbits. The data were produced using the numerical method of Ref.~\cite{Shah:2012gu}. Parenthetical figures are estimated numerical error bars on the last displayed decimals; 
e.g., $1.245\,568(2)$ stands for $1.245\,568\pm 0.000\,002$.
}
\label{table:data}
\end{table}

Table \ref{table:data} and Fig.~\ref{fig:ISCEO-shift} show numerical results for $C_{\Omega}(q)$. Method (i) gives $C_{\Omega}(0) = 1.251\,015\,39 \pm 4\times 10^{-8}$, a value consistent with (and as accurate as) the most accurate $q=0$ result currently available \cite{Akcay:2012ea}.
We notice that $C_{\Omega}(q)$ reaches minimum at $q \simeq 0.3$ and is 
positive for all values in our sample ($|q|\leqslant 0.9$).

The lower panel in Fig.\ \ref{fig:ISCEO-shift} tests three earlier predictions for $C_{\Omega}(q)$ against our ``exact'' self-force result: two based on the PN approximation, at 3 PN order using a stability analysis of the equations of motion \cite{Favata:2010ic} and at 3.5PN order using a MECO condition with a certain modified binding energy function \cite{Tessmer:2012xr}, and one arising from the EOB model \cite{Barausse:2009xi} with the method described in Refs. \cite{Buonanno:2000ef,Hinderer:2013uwa}.
This comparison illustrates the discriminative power of our result as a new benchmark in the strong-field regime. We expect that our results will be used in refining semianalytical models of inspiralling binaries over the full range of mass ratios and spins~\cite{Taracchini:2013rva,Hinder:2013oqa}.

\acknowledgments
We wish to thank to Ryuichi Fujita, Charalampos Markakis 
and Norichika Sago for their helpful feedback.
We also would like to thank Enrico Barausse for providing us (on request) ISCO
frequency shift data from the EOB model of Ref.~\cite{Barausse:2009xi}.
S.I. acknowledges the support of the Grant-in-Aid for JSPS Fellows 
(No.\ 24-4281), and is grateful for the warm hospitality 
of all members in the gravity group at University of Southampton, 
where this work was finalized.
L.B. and A.G.S. acknowledge funding from the European Research Council under the European Union's Seventh Framework Programme FP7/2007-2013/ERC Grant No.~304978, and L.B. acknowledges additional support from STFC through Grant No.\ PP/E001025/1. 
S.D. thanks Barry Wardell for many helpful conversations and shared code.
A.L.T. gratefully acknowledges the hospitality of the Yukawa Institute for Theoretical Physics, where part of this work was completed.
H.N. and T.T. acknowledge the support of the Grant-in-Aid for Scientific Research on Innovative Area (No.\ 24103006).
T.T. is also supported by the Grant-in-Aid for Scientific Research (No.\ 26287044) and for Scientific Research on Innovative Area (No.\ 24103001, and No.\ 24103006).
N.W.'s work was supported by the Irish Research Council, which is funded under
the National Development Plan for Ireland.


\end{document}